\pdfoutput=1

\documentclass[runningheads,a4paper]{llncs}

\usepackage{cmap}
\usepackage[T1]{fontenc}

\usepackage{graphicx}
\usepackage{booktabs}
\usepackage{listings}
\usepackage{algorithmicx}
\usepackage{amsmath,amsfonts,amssymb}
\usepackage{mathtools}
\DeclarePairedDelimiter{\ceil}{\lceil}{\rceil}
\usepackage{amsfonts}
\usepackage{amssymb}
\usepackage[breakall]{truncate}
\usepackage{caption}
\usepackage{subcaption}
\usepackage{multirow}
\usepackage{qtree}
\usepackage{tabularx}

\usepackage[ngerman,english]{babel}
\addto\extrasenglish{\languageshorthands{ngerman}\useshorthands{"}}

\usepackage[%
rm={oldstyle=false,proportional=true},%
sf={oldstyle=false,proportional=true},%
tt={oldstyle=false,proportional=true,variable=true},%
qt=false%
]{cfr-lm}
\usepackage[math]{blindtext}

\usepackage{cite}

\usepackage{paralist}

\usepackage[utf8]{inputenc}
\usepackage{csquotes} 

\usepackage{microtype}

\usepackage[hyphens]{url}

\def\UrlOrds{\do\*\do\-\do\~\do\'\do\"\do\-\do\/}

\makeatletter
\g@addto@macro{\UrlBreaks}{\UrlOrds\do\/}
\makeatother

\usepackage[table]{xcolor}
\usepackage{tikz}
\usetikzlibrary{shapes,arrows,calc,plotmarks}

\usepackage{siunitx}
\usepackage{xspace} 
\usepackage{footnote}

\usepackage[hidelinks]{hyperref}
\hypersetup{breaklinks=true}

\usepackage[capitalise,nameinlink]{cleveref}

\crefname{section}{Sect.}{Sect.}
\Crefname{section}{Section}{Sections}

\makesavenoteenv{tabular}
\makesavenoteenv{table}

\DeclareCaptionType{mycapequ}[][List of equations]
\newcolumntype{L}[1]{>{\raggedright\let\newline\\\arraybackslash\hspace{0pt}}m{#1}}

\newcommand{\eg}{e.\,g.,\ }
\newcommand{\ie}{i.\,e.,\ }

\DeclareFontFamily{U}{MnSymbolC}{}
\DeclareSymbolFont{MnSyC}{U}{MnSymbolC}{m}{n}
\DeclareFontShape{U}{MnSymbolC}{m}{n}{
    <-6>  MnSymbolC5
   <6-7>  MnSymbolC6
   <7-8>  MnSymbolC7
   <8-9>  MnSymbolC8
   <9-10> MnSymbolC9
  <10-12> MnSymbolC10
  <12->   MnSymbolC12%
}{}
\DeclareMathSymbol{\powerset}{\mathord}{MnSyC}{180}

\newcommand\hmm[1]{\ifnum\ifhmode\spacefactor\else2000\fi>1000 \uppercase{#1}\else#1\fi}

\newcommand{\EP}[0]{Ethereum platform\xspace}
\newcommand{\ET}[0]{Ethereum\xspace}

\newcommand{\EA}[0]{et al.\xspace}
\newcommand{\ETC}[0]{etc.\@\xspace}

\newcommand{\ERC}[0]{ERC-20\@\xspace}

\newcommand{\TOKS}[0]{token systems\@\xspace}
\newcommand{\TOK}[0]{token system\@\xspace}

\newcommand{\hexit}[1]{\texttt{0x#1}}

\newcommand{\airdropper}{\hmm{d}istributor\xspace}
\newcommand{\airdroppers}{\hmm{d}istributors\xspace}
\newcommand{\airdroppee}{\hmm{r}ecipient\xspace}
\newcommand{\airdroppees}{\hmm{r}ecipients\xspace}
\newcommand{\tx}{\textbf{\texttt{tx:}}\xspace}
\newcommand{\call}{\textbf{\texttt{call:}}\xspace}

\lstdefinelanguage{Solidity}{
  keywords={typeof, enum, new, external, true, false, catch, function, event, modifier, internal, private, sender, require, revert, return, null, catch, public, returns, switch, var, if, in, while, do, else, case, break, contract, throw, value, data, sig, gas, gasprice, origin, coinbase, difficulty, blockhash, timestamp, number, gaslimit, assert, require, revert, addmod, mulmod, selfdestruct, sha3, keccak256, ecrecover, sha256, ripemd160, dataCopy, pragma, solidity, constant, for, continue, ?, :, storage, memory},
  ndkeywords={class, export, implements, import, this, payable, view, pure, now, msg, block, wei, day, call, tx, send, delegatecall, callcode, address, uint, bytes2, bytes, fixedMxN, bytes32, void, bytesN, bool, bytes20, uint8, uint16, int16, uint248, int, uint256, int256, bytes4, uint32, int8, bytes1, byte, bytes32, mapping, string, struct},   
  basicstyle=\scriptsize\ttfamily,
  keywordstyle=\color{blue},
  ndkeywordstyle=\color{orange},
  commentstyle=\color{cyan},
  numberstyle=\color{gray},
  stringstyle=\color{purple},
  breakatwhitespace=false,         
  breaklines=true,                 
  captionpos=b,                    
  keepspaces=true,                 
  numbers=left,                    
  numbersep=5pt,                  
  showspaces=false,                
  showstringspaces=false,
  showtabs=false,                  
  tabsize=2,
  comment=[l]{//},
}

\lstdefinelanguage{EthereumBytecode}{
    keywords={PUSH1, CALLDATALOAD, PUSH4, EQ,PUSH1, EQ, JUMPI, JUMPDEST},
    ndkeywords={0x4, 0xa9059cbb, 0x20, 0x4},   
    basicstyle=\scriptsize\ttfamily,
    keywordstyle=\color{blue},
    ndkeywordstyle=\color{orange},
    commentstyle=\color{cyan},
    numberstyle=\color{black},
    stringstyle=\color{black},
    breakatwhitespace=false,         
    breaklines=true,                 
    captionpos=b,                    
    keepspaces=true,                 
    numbers=none,                    
    numbersep=5pt,                  
    showspaces=false,                
    showstringspaces=false,
    showtabs=false,                  
    tabsize=2,
    comment=[l]{//},
}

\lstdefinelanguage{JavaScript}{
  keywords={typeof, new, true, false, catch, function, return, null, catch, switch, var, if, in, while, do, else, case, break, console, =>},
  ndkeywords={class, export, boolean, throw, implements, import, this},
  basicstyle=\scriptsize,
  keywordstyle=\color{blue},
  commentstyle=\color{cyan},
  numberstyle=\color{gray},
  stringstyle=\color{purple},
  breakatwhitespace=false,         
  breaklines=true,                 
  captionpos=b,                    
  keepspaces=true,                 
  numbers=none,                    
  numbersep=5pt,                  
  showspaces=false,                
  showstringspaces=false,
  showtabs=false,                  
  tabsize=2,
  comment=[l]{//},
}

\usepackage{chngcntr}
\AtBeginDocument{\counterwithout{lstlisting}{chapter}}

\newcommand{\soltext}[1]{{\lstinline[columns=fixed, language=Solidity, basicstyle=\ttfamily]$#1$}}
\newcommand{\sol}[1]{{\lstinline[columns=fixed, language=Solidity]$#1$}}

\newcommand{\approachlbl}[2][black]{\textsf{\textcolor{#1}{\scriptsize#2}}}

\begin{document}
\title{The Operational Cost of Ethereum Airdrops}

\author{Michael Fr"owis \and Rainer B"ohme}
\institute{Department of Computer Science, Universit\"at Innsbruck, Austria}

%




            
\maketitle

\begin{abstract}
Efficient transfers to many recipients present a host of issues on \ET. First, accounts are identified by long and incompressible constants. Second, these constants have to be stored and communicated for each payment. Third, the standard interface for token transfers does not support lists of recipients, adding repeated communication to the overhead. Since \ET charges resource usage, even small optimizations translate to cost savings.
Airdrops, a popular marketing tool used to boost coin uptake, present a relevant example for the value of optimizing bulk transfers. Therefore, we review technical solutions for airdrops of \ET-based tokens, discuss features and prerequisites, and compare the operational costs by simulating 35 scenarios. We find that cost savings of factor two are possible, but require specific provisions in the smart contract implementing the \TOK. Pull-based approaches, which use on-chain interaction with the \airdroppees, promise moderate savings for the \airdropper while imposing a disproportional cost on each \airdroppee.
Total costs are broadly linear in the number of \airdroppees independent of the technical approach.
We publish the code of the simulation framework for reproducibility, to support future airdrop decisions, and to benchmark innovative bulk payment solutions.
\end{abstract}


\section{Introduction}
\label{sec:intro}

Fungible virtual assets, such as cryptocoins and tokens residing on a blockchain, are network goods: their value lies in enabling exchange. A coin is worthless if nobody else uses or accepts it. Its value grows quadratically in the number of users, according to Metcalfe's law; and still super-linear under more conservative theories~\cite{Briscoe2006}. As a result, new coins have to reach a critical mass until positive feedback sustains rapid growth~\cite{Boehme2013-IAB}. 

This observation is taken to heart in the marketing of new coins. The community has adopted the term \emph{airdrop} for the subsidized (often free) provision of new coins to selected lead users, typically holders of competing coins, with the intention to raise popularity and reach critical mass. Similar strategies are well understood in the economics~\cite{Katz1994} and marketing literature~\cite{Hill2006}. Whether and under which conditions airdrops are successful for cryptocoins and tokens are empirical questions that future work should tackle. Here, we study the operational costs of airdrops on \ET, the most popular platform for \TOKS. 

Airdrops incur costs in the form of transactions fees paid to miners, which are shared between the initiator of the airdrop (often the developer or maintainer of a token, henceforth \emph{\airdropper}) and the \emph{\airdroppees}, depending on the technical approach chosen by the \airdropper. The platform charges fees for instructions, space on the blockchain, and the size of the state information. The costs are not negligible because every recipient identifier contains cryptographic material with high entropy that must be communicated in the airdrop. Typically, the identifiers are included in the transaction payload and thus occupy space on the blockchain. Another difficulty faced by \ET airdrops is the lack of a bulk transaction method in the popular \ERC standard~\cite{ercstd} for fungible tokens. This adds overhead due to repeated communication.



We have observed several solutions and workarounds to these problems in the wild, and synthesize our findings into the---to the best of our knowledge---first systematic overview on the technology behind airdrops on the \EP. We implement selected techniques in model smart contracts and measure their cost and resource consumption as a function of the number of \airdroppees by executing the contracts in a simulated \ET node.


The rest of the paper is organized as follows. 
Sections~\ref{sec:tech} presents technical options for carrying out airdrops on \ET, including a discussion of the relevant parameters and necessary prerequisites. We distinguish push and pull approaches, internal and external batching, and revisit pooled payments.
Section~\ref{sec:operational_cost} presents the cost estimates from the simulation study in units of \ET's internal fee model (gas), the best level of analysis for comparison between options. 
Section~\ref{sec:discussion} interprets the main findings in units of fiat currency (USD), the level of analysis that matters for business decisions. 
Section~\ref{sec:related_work} connects to relevant related work, before we give an outlook and conclude in Section~\ref{sec:conclusion}.
Technical details of the simulation framework are placed in two appendices.


\section{Technical Aspects of Ethereum Airdrops}
\label{sec:tech}
This section gives an overview about technical considerations when conducting airdrops on \ET. 
We briefly discuss parameters of an airdrop chosen by the business side, then explain shortcomings of the default token transfer interface and the resulting technical workarounds.
We apply the lens of operational costs, which in case of the \EP translates to estimating transaction fees in units of gas, the most comparable metric. 

\subsection{Parameters of an Airdrop}
Before carrying out an airdrop, a couple of parameters need to be decided.
One of the first things to decide on is \emph{who} shall receive tokens.
This is often done by a simple sign-up system (using \eg Telegram, web forms, \ETC), or by defining a measure of relevance on existing addresses to select the set of \airdroppees. 
Companies like \emph{Bounty One}\footnote{\url{https://bountyone.io/airdrops}, [Online; accessed 18 Jun 2019].} offer matching between airdrop- \airdroppers and  \airdroppees as a service.
The rationale behind this is simple: you prefer to hand out tokens to active users participating in the ecosystem, instead of sending them to inactive accounts. 
For example, the banking startup \emph{OmiseGO} conducted one of the early \ET airdrops~\cite{OMISEDetails}. They used account balance as an activity indicator and simply handed out tokens to every address holding more than \num{0.1} Ether (ETH) at block height \num{3988888}. This airdrop serves as good running example because all code, including a documentation of the rationales behind design decisions, is publicly available~\cite{OmiseGOAD}. 
The threshold of \num{0.1} ETH is a very simple metric. It does not consider essential aspects, such as: are those accounts still active and able to use the tokens?\footnote{This depends on the account type and  state. For example, disabled contracts or contracts that are not programmed to transact with token systems will never be able to use the funds. This is also noted in~\cite{OmiseGOAD}.} 

Two other important parameters of an airdrop are the number of tokens to be dropped and their distribution over \airdroppees. 
For example, OmiseGO dropped \num{5}\% of the total supply of OMG tokens.
The distribution between \airdroppees can be uniform or depend on properties of the \airdroppee.
OmiseGO allocated tokens to  \airdroppees proportional to their ETH balances at a specified point in time.\footnote{See \url{https://github.com/omisego/airdrop/blob/master/processor.py}, line 77}.


Finally, the technical approach of how to transfer the tokens to the recipients needs to be defined.
This involves choosing the software implementation to distribute tokens in bulk.
This decision heavily depends on the existing infrastructure of the underlying \TOK. 

In summary, the main decisions to be taken are:
\begin{itemize}
  \item Who receives tokens (based on what metric)?\hfill \emph{(\airdroppee selection)}
  \item How many tokens per \airdroppee? \hfill \emph{(distribution)}
  \item Which technical approach to use for distributing the tokens? \hfill \emph{(implementation)}
\end{itemize}

Although \airdroppee selection and the token distribution strategy involve technical aspects, they are mainly driven by business considerations. Those are out of the scope of this work. Here, we focus on technical implementation options and their associated cost. Our results are an important input to the business decision because they quantify the operational cost of an airdrop.

\subsection{Shortcomings of Vanilla \ERC{} Airdrops}
\label{sec:erc20_short}

\begin{figure}[t]
\centering
  \begin{tikzpicture}[node distance=2cm,auto,>=stealth, x=5cm,y=4.5cm]

    \coordinate (a) at (0,0);
    \coordinate (b) at (0,1);
    \coordinate (c) at (1,0);
    \coordinate (d) at (1,1);

    \node[rectangle, line width=0] at (.5, .4)  (dot)     {\vdots};

    \draw (a) -- (b)node[pos=1.1, above, text width=1.75cm, align=center]{\airdropper EOA}
          (c) -- (d)node[pos=1.1, above, text width=1.75cm, align=center]{\ERC \TOK};

    \draw[-stealth] ($(a)!0.75!(b)$) -- node[above,midway]{\tx \sol{transfer(recipient1, 500)}*}($(c)!0.75!(d)$);
    \draw[-stealth] ($(a)!0.65!(b)$) -- node[above,midway]{\tx \sol{transfer(recipient2, 500)}*}($(c)!0.65!(d)$);
    \draw[-stealth] ($(a)!0.55!(b)$) -- node[above,midway]{\tx \sol{transfer(recipient3, 500)}*}($(c)!0.55!(d)$);

    \node[rectangle, line width=0] at (.5, .4)  (dot)     {\vdots};

    \draw[-stealth] ($(a)!0.15!(b)$) -- node[above,midway]{\tx \sol{transfer(recipient1000, 500)}*}($(c)!0.15!(d)$);
    
  \end{tikzpicture}
  \caption{Na\"ive push-style airdrop. One transaction per recipient from \airdropper to \TOK. The * indicates that the method is part of the \ERC standard.}
  \label{fig:naive_airdrop}
\end{figure}
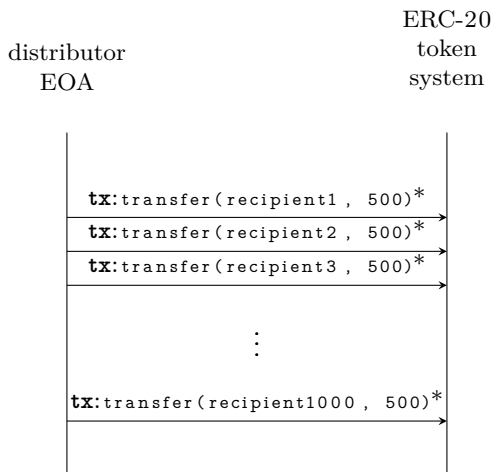

\ERC\cite{ercstd} is the most prominent standard on the \EP today. It defines an API for token systems that enables:
(1) the encoding of token properties,\footnote{Functions: \sol{symbol}, \sol{name}, \sol{decimals}, \sol{totalSupply}}
(2) access to the balances of owners,\footnote{Functions: \sol{balanceOf}} and
(3) the transfer of tokens between accounts.\footnote{Functions: \sol{transfer}, \sol{transferFrom}, \sol{approve}, \sol{allowance}}
Moreover, \ERC defines logging and event notification.\footnote{Events: \sol{Approve} and \sol{Transfer}}

The first \ERC \TOKS emerged already in late 2015.\footnote{The first \ERC token. Block: \num{490326},\\ Address: \hexit{Eff6425659825E22a3cb00d468E769f038166ae6}}
Airdrops, on the other hand, are a more recent phenomenon starting to gain traction in early 2018.
As a consequence, the \ERC API does not include a batch transfer method to directly transfer tokens to \emph{multiple} recipients in one transaction. The lack of this functionality in legacy token systems makes airdrops more expensive. The immutability of contracts, which is often a desired feature, prevents legacy token systems from adding batch capabilities after deployment. 

Hence, implementing an airdrop in vanilla \ERC proceeds as shown in Figure~\ref{fig:naive_airdrop}. The \airdropper uses an externally owned account (EOA) in order to send one transaction for every \airdroppee. Each transaction invokes the \soltext{transfer} method of the token system in order to update its internal state. The fixed cost per transaction is \num{21000} units of gas, which are not recoverable and add to the overhead of this approach.
To avoid issuing one full transaction per airdrop \airdroppee, the community developed several optimizations to reduce cost and avoid network congestion.



\subsection{Optimizations}
\label{sec:opti}

We distinguish three avenues for improvements. Transaction batching helps to reduce communication costs. The pull approach shifts part of the burden to the \airdroppee and potentially conserves tokens and fees from \airdroppees who do not collect their share. Off-chain approval saves cost by avoiding to store the list of account identifiers on the blockchain. We discuss each of these avenues in the following subsections.


\subsubsection{Transaction Batching}\hfill
\label{sssec:batch}

The easiest way to save cost is by removing the overhead of issuing one transaction for each airdrop \airdroppee. A cheaper alternative to transactions are \emph{message calls} (also known as \emph{internal transactions}) invoked by contracts. The difference in fixed cost (without payload) is substantial: a transaction costs \num{21000} units of gas, compared to \num{700}  for a message call. Message calls to the same contract are yet an order of magnitude cheaper and cost about \num{10} units of gas. Message calls are generated in a loop over the list of \airdroppees, which is given as an argument to a single initial transaction.

\begin{figure}[ht]
\centering
  \begin{tikzpicture}[node distance=2cm,auto,>=stealth, x=5cm,y=5cm]

    \coordinate (a) at (0,0);
    \coordinate (b) at (0,1);

    \coordinate (c) at (1.1,0);
    \coordinate (d) at (1.1,1);

    \coordinate (e) at (2,0);
    \coordinate (f) at (2,1);

    \node[rectangle, line width=0] at (.55, .2)  (dot)     {\vdots};

    \draw (a) -- (b)node[pos=1.1, above, text width=1.75cm, align=center]{\airdropper EOA}
          (c) -- (d)node[pos=1.1, above, text width=1.75cm, align=center]{batching contract}
          (e) -- (f)node[pos=1.1, above, text width=1.75cm, align=center]{\ERC \TOKS};

    \draw[-stealth] ($(a)!0.75!(b)$) -- node[above,midway, text width=5cm, align=center]{\tx\\\sol{transferMany([r1, r2, ...], 500)}}($(c)!0.75!(d)$);
    \draw[-stealth] ($(a)!0.3!(b)$) -- node[above,midway, text width=6cm, align=center]{\tx\\\sol{transferMany([r10, r11, ...], 500)}}($(c)!0.3!(d)$);

    \draw[-stealth] ($(c)!0.75!(d)$) -- node[above,midway]{\call~\sol{transfer(r1, 500)}*}($(e)!0.75!(f)$);
    \node[rectangle, line width=0] at (1.8, .65)  (dot)     {\vdots};
    \draw[-stealth] ($(c)!0.75!(d)$) -- node[below,midway,rotate=-18]{\call~\sol{transfer(r10, 500)}*}($(e)!0.45!(f)$);

    \draw[-stealth] ($(c)!0.3!(d)$) -- node[above,midway]{\call~\sol{transfer(r10, 500)}*}($(e)!0.3!(f)$);
    \node[rectangle, line width=0] at (1.8, .2)  (dot)     {\vdots};
    \draw[-stealth] ($(c)!0.3!(d)$) -- node[below,midway,rotate=-18]{\call~\sol{transfer(r20, 500)}*}($(e)!0.0!(f)$);
    
  \end{tikzpicture}
  \caption{External batching, push-style airdrop. The \TOK has \textbf{no} batching capabilities. Batching is done by an external contract (either own or service\protect\footnotemark). Note the calls instead of transactions.}
  \label{fig:external_push}
\end{figure}
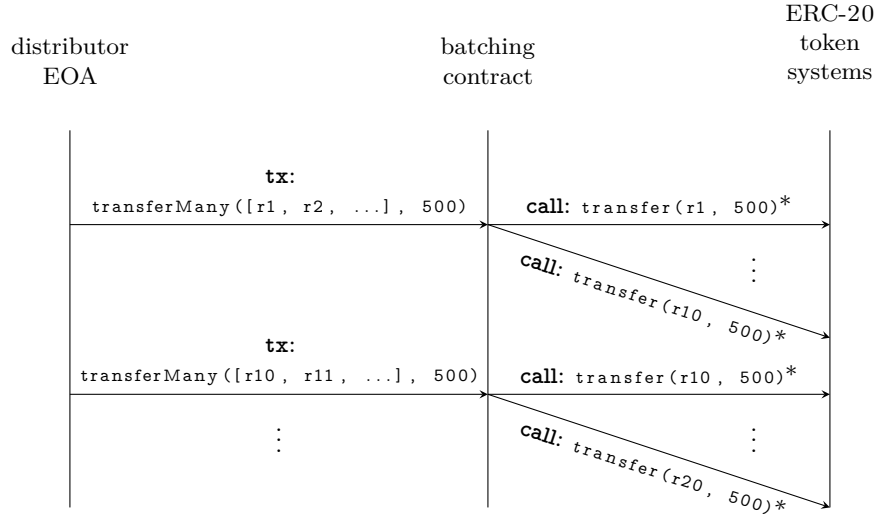

Let $n$ be the number of \airdroppees, then the cost savings $s$ are
\begin{equation}
	\label{eq:save-ext}
	s_1 = n\cdot21000 - (n\cdot 700 + 21000),
\end{equation}
if the loop is implemented in a different contract than the \TOK. This method is called \emph{external batching} and visualized in Figure~\ref{fig:external_push}.
To give an intuition for the source of the savings, recall that the batch is authorized by a single signature as compared to to one signature per transaction in the na\"ive approach.

Even higher savings of
\begin{equation}
	\label{eq:save-int}
	s_2 = n\cdot21000 - (n\cdot 10 + 21000)
\end{equation}
are possible if the loop is implemented directly in the \TOK contract (\emph{internal batching}). The difference between external and internal batching can be explained by the penalty of fetching new code from disk, which applies $n$ times in the case of external batching. However, changing the \TOKS's contract may either be impossible because it is already deployed immutably, or not desired in the fear of introducing new bugs or losing investor trust. 

The savings are upper bounds that are only achievable if all $n$ identifiers fit into one transaction. The size of transactions is restricted by the block gas limit.\footnote{Gas limit at the time of writing is \num{8000029} in block number \num{8014738}} Larger \airdroppee lists must be split into several batches, each incurring the fixed cost of one transaction.

\footnotetext{For example, \url{https://multisender.app/}, [Online; accessed 25 Jun 2019].}







\subsubsection{The Pull Approach}\hfill
\label{sec:pull}

As already mentioned, distributing tokens to inactive accounts or to \airdroppees not interested in the token has little to no value to the \airdropper.
This problem can be addressed during \airdroppee selection, by 
\begin{itemize}
\item evaluating appropriate technical indicators,
\item collecting information provided in sign-up, or
\item using the specialized services in the ecosystem who administer panels of potential \airdroppees.\footnote{\url{https://bountyone.io/airdrops}, [Online; accessed 18 Jun 2019].}
\end{itemize}
We are not aware of any literature on the effectiveness of these options and consider it out of our scope.

A more technical approach is to condition the token transfer on on-chain user interaction. 
This type of \emph{pull-style airdrop} can be implemented using the \soltext{approve} function of \ERC.
Instead of directly transferring tokens to the \airdroppees during the airdrop, the \airdropper gives the \airdroppee the right to withdraw the airdropped amount. 
The \airdropper may specify a deadline for the withdrawal and reclaim the remaining tokens thereafter.

Arguably, the additional effort and cost for the \airdroppee ensures that the distribution is more targeted. 

\begin{figure}[h]
\centering
  \begin{tikzpicture}[node distance=2cm,auto,>=stealth, x=5cm,y=4.5cm]

    \coordinate (a) at (0,0);
    \coordinate (b) at (0,1);

    \coordinate (c) at (1,0);
    \coordinate (d) at (1,1);

    \coordinate (e) at (2,0);
    \coordinate (f) at (2,1);

    \coordinate (g) at (2.05,-0.1);
    \coordinate (h) at (2.05,0.9);

    \coordinate (i) at (2.1,-0.2);
    \coordinate (j) at (2.1,0.8);

    \node[rectangle, line width=0, rotate=-40] at (2.175, .35)  (dot)     {\dots};

    \coordinate (k) at (2.24,-0.4);
    \coordinate (l) at (2.24,0.6);

    \draw (a) -- (b)node[pos=1.1, above, text width=1.75cm, align=center]{\airdropper EOA \\ \sol{(Aird)}}
          (c) -- (d)node[pos=1.1, above, text width=1.75cm, align=center]{\ERC \TOK}
          (e) -- (f)node[pos=1.1, above, text width=1.75cm, align=center]{r1}
          (g) -- (h)node[pos=1.1, above, text width=1.75cm, align=center]{r2}
          (i) -- (j)node[pos=1.1, above, text width=1.75cm, align=center]{r3}
          (k) -- (l)node[pos=1.1, above, text width=1.75cm, align=center]{r1000};

    \draw[-stealth] ($(a)!0.9!(b)$) -- node[above,midway, text width=5cm, align=center]{\tx\\\sol{approveMany([r1, r2, ...], 500)}}($(c)!0.9!(d)$);

    \node[rectangle, line width=0] at (.5, .8)  (dot)     {\vdots};

    \draw[stealth-, dashed] ($(c)!0.75!(d)$) -- node[above, midway, text width=5cm, align=center]{\tx\\\sol{transferFrom(Aird, r1, 500)}*}($(e)!0.75!(f)$);

    \draw[-stealth] ($(a)!0.5!(b)$) -- node[above,midway, text width=5cm, align=center]{\tx\\\sol{approveMany([r1000, ...], 500)}}($(c)!0.5!(d)$);
    
    \draw[stealth-,  dashed] ($(c)!0.2!(d)$) -- node[above,midway, xshift=-0.6cm, text width=5.13cm, align=center]{\tx\\\sol{transferFrom(Aird, r1000, 500)}*}($(k)!0.6!(l)$);
  \end{tikzpicture}
  \caption{Pull-style airdrop with non-standard \soltext{approveMany} for internal batching. Recipients have to interact with the \ERC contract to finally receive the funds. The \airdropper-side can be batched. Recipients must use individual transactions. Dashed lines mean \airdroppee pays.}
  \label{fig:Pull}
\end{figure}
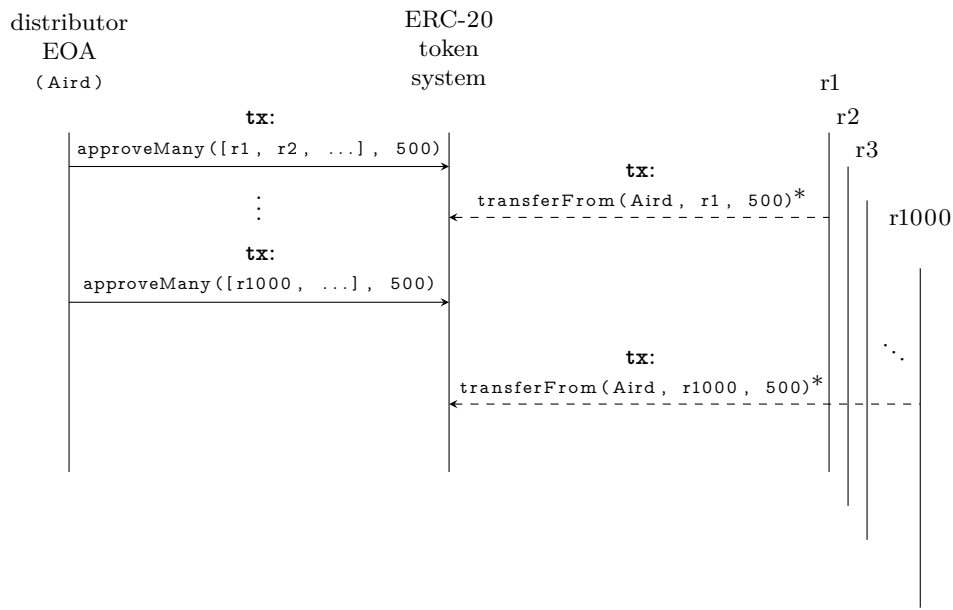

The pull approach has a couple of downsides. 
First, the cost for the \airdroppee and the \airdropper are significant. Both sides pay about as much as the \airdropper pays for a push-style airdrop. This even holds when the \airdropper approves in batches as described in Section~\ref{sssec:batch} (see Figure~\ref{fig:Pull}). Second, a known front-running attack against the \ERC\ \soltext{approve} logic requires the \airdropper to set all allowances to zero before updating them with new values\cite{approvalVulnerable,ercstd}. 
This approximately doubles his cost.
Third, many existing \TOKS do not implement the \soltext{approve} method and therefore cannot use the pull approach~\cite{frowis2018detecting}.
Lastly, recovering the unclaimed tokens after the deadline costs about as much as transferring them.

\subsubsection{Off-Chain Approval}\hfill
\label{sec:pooledpay}

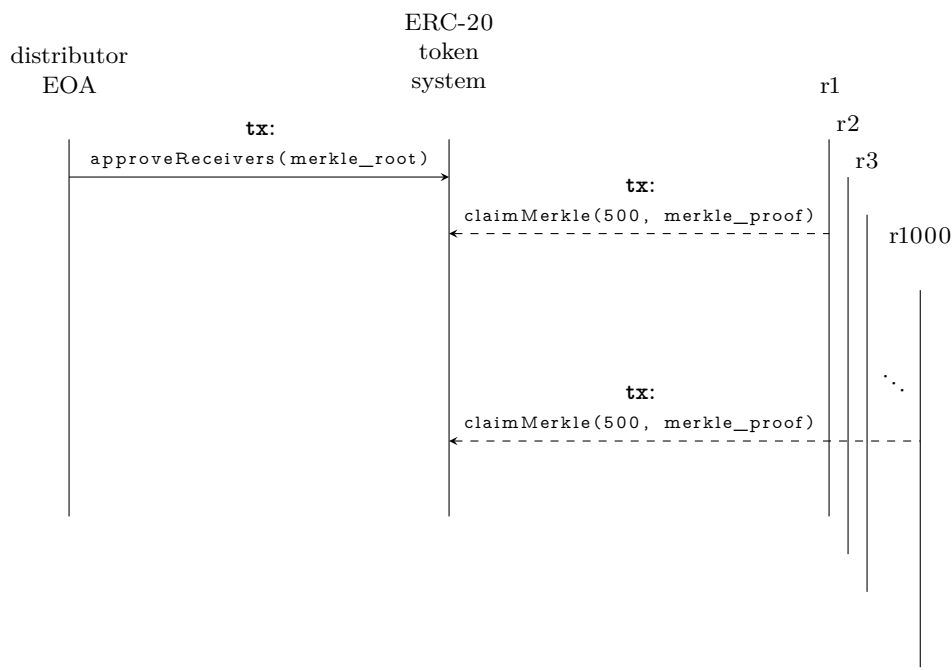
\begin{figure}[h]
\centering
  \begin{tikzpicture}[node distance=2cm,auto,>=stealth, x=5cm,y=5cm]

    \coordinate (a) at (0,0);
    \coordinate (b) at (0,1);

    \coordinate (c) at (1,0);
    \coordinate (d) at (1,1);

    \coordinate (e) at (2,0);
    \coordinate (f) at (2,1);

    \coordinate (g) at (2.05,-0.1);
    \coordinate (h) at (2.05,0.9);

    \coordinate (i) at (2.1,-0.2);
    \coordinate (j) at (2.1,0.8);

    \node[rectangle, line width=0, rotate=-40] at (2.175, .35)  (dot)     {\dots};

    \coordinate (k) at (2.24,-0.4);
    \coordinate (l) at (2.24,0.6);

    \draw (a) -- (b)node[pos=1.1, above, text width=1.75cm, align=center]{\airdropper EOA}
          (c) -- (d)node[pos=1.1, above, text width=1.75cm, align=center]{\ERC \TOK}
          (e) -- (f)node[pos=1.1, above, text width=1.75cm, align=center]{r1}
          (g) -- (h)node[pos=1.1, above, text width=1.75cm, align=center]{r2}
          (i) -- (j)node[pos=1.1, above, text width=1.75cm, align=center]{r3}
          (k) -- (l)node[pos=1.1, above, text width=1.75cm, align=center]{r1000};

    \draw[-stealth] ($(a)!0.9!(b)$) -- node[above,midway, text width=5cm, align=center]{\tx\\\sol{approveReceivers(merkle_root)}}($(c)!0.9!(d)$);

    \draw[stealth-, dashed] ($(c)!0.75!(d)$) -- node[above, midway, text width=5cm, align=center]{\tx\\\sol{claimMerkle(500, merkle_proof)}}($(e)!0.75!(f)$);
    \draw[stealth-, dashed] ($(c)!0.2!(d)$) -- node[above,midway, xshift=-0.6cm, text width=5.13cm, align=center]{\tx\\\sol{claimMerkle(500, merkle_proof)}}($(k)!0.6!(l)$);
    
  \end{tikzpicture}
  \caption{Pooled payments: non-standard, internal, pull-style airdrop. Cost is constant for \airdropper. List of \airdroppees is public. }
  \label{fig:merkle_mine}
\end{figure}

The \ET community has realized that storing every \airdroppee address on the chain leads to network congestion as well as to high cost. One approach that does not require to store all \airdroppee addresses on-chain are \emph{pooled payments}~\cite{PooledPayments}.
They are inspired by \emph{Merkle mine}\cite{MerkleMine}, an approach developed for \TOKS to define the initial allocation to a large number of owners.

Pooled payments resemble pull-style airdrops, as the \airdropper approves the \airdroppee to withdraw a certain amount of tokens through a transaction. 
But pooled payments do not store the entire approval on-chain. 
Instead, the \airdropper encodes the list of \airdroppees with denominations in a Merkle tree~\cite{merkle1987digital}, where leafs are concatenations of \airdroppee addresses and amounts. 
The approving contract has to store the root hash of the Merkle tree only (see Fig.~\ref{fig:merkle_mine}, \soltext{approveReceivers(merkle_root)}). 
The list of \airdroppees is published off-chain.
To claim funds, every \airdroppee needs the list and computes the Merkle tree as well as a Merkle proof for his entry. 
The \airdroppee then sends a transaction to the contract with his address (implicit by the signature on the transaction), the amount, and the Merkle proof (see Fig.~\ref{fig:merkle_mine}, \soltext{claimMerkle(500, merkle_proof)}).
This allows the contract to compute the hash of the leaf node from the message sender (signature) and the amount in the argument.  Using the Merkle proof, it verifies the correctness of the claim, checks its freshness, and transfers the funds. 
To prevent double-claiming, the contract must store a record of this withdrawal. Several methods exist to keep this as compact and cost-efficient as possible.

Unlike for normal pull payments, the \airdropper has constant cost independent of the number of recipients. 
Most of the airdrop cost is shifted to the \airdroppee.\footnote{This accounts to: one transaction per \airdroppee, Merkle proof verification, and storage of withdrawal record.} 
 
It is also worth mentioning that pooled payments based on off-chain approvals have some usability issues that may delay their adoption for airdrops.
Both \airdroppees and \airdroppers need tools to do off-chain computation (Merkle tree, proofs) and data retrieval (list of \airdroppees).
We are aware of one business that seems to bet on the adoption of this approach.\footnote{The claim of constant \airdropper cost in the Coinstantine whitepaper indicates the use of pooled payments. See \url{https://www.coinstantine.io/}, [Online; accessed 22 Jun 2019].}


\subsection{Miscellaneous Aspects}

Optimizing the airdrop strategy and the code involved in the airdrop workflow are not the only things to consider when doing airdrops.

\textbf{Gas Token:}
Gas Token\footnote{\url{https://gastoken.io/}, [Online; accessed 21 Jun 2019]} provides a way to pre-acquire gas in periods when gas is cheap.
Those gas tokens can be \textquote{redeemed} when the gas price is high.
The gas token mechanism exploits the fact that \ET refunds gas when storage resources are freed.
To include this mechanism in airdrops, the functionality to redeem gas tokens must be built into either the \TOK or the batching contract.
Gas tokens are already used in other areas, such as arbitrage bots~\cite{daian2019flash}.

\textbf{Systemic Risk:}
Airdrops can be a systemic risk for the \EP, if not used carefully, since they use large amounts of resources (gas).
Coindesk\cite{AirdropGasCrisis} reports that the uptake of airdrops in conjunction with questionable incentives set by the exchange FCOIN led to substantial network congestion, gas price increases, and wasted resources.
Reportedly, OmiseGO also considered the impact of its airdrop on the network and decided to limit their batches such that they never use more than 50\% of the block gas limit\cite{OmiseGOAD}. We adopt the idea of such a limit for the simulations in the following.


\section{Cost Estimates}
\label{sec:operational_cost}

In the following we compare simulated costs of different airdrop techniques. The simulation gives us valid estimates of total cost, which puts the back-of-the-envelope calculations of savings in Eqs.~(\ref{eq:save-ext}) and (\ref{eq:save-int}) into perspective.
The simulation framework, the constants used, a complete list of scenarios, and the code can be found in Appendix~\ref{sec:simulation_setup} and \ref{sec:contracts}. 
It can serve as starting point for facts-based airdrop decisions and to benchmark new solutions.

All 35 scenarios are combinations of the approaches discussed in Section~\ref{sec:opti}. 
Table~\ref{tbl:labels} resolves the labels used in the figures below.
\begin{table}[h]
\centering
\caption{Approach labels and descriptions.}
\begin{tabularx}{\linewidth}{lX@{}}
  \toprule
  \textbf{Label} & \textbf{Description} \\
  \midrule
  \textbf{\textsf{\scriptsize NAIVE:}} & No batching is applied. One transaction per \airdroppee. \\
  \textbf{\textsf{\scriptsize PUSH:}} & Push-style airdrop as discussed in Section~\ref{sec:opti}. \\
  \textbf{\textsf{\scriptsize PULL:}} & Pull-style airdrop as discussed in Section~\ref{sec:opti}. \\
  \textbf{\textsf{\scriptsize EXTERNAL\_BATCH:}} & External batching as discussed in Section~\ref{sec:opti}. \\
  \textbf{\textsf{\scriptsize INTERNAL\_BATCH:}} & Internal batching as discussed in Section~\ref{sec:opti}. \\
  \textbf{\textsf{\scriptsize UNIFORM:}} & One amount per batch. Otherwise $n$ different amounts are sent. \\
  \textbf{\textsf{\scriptsize RECIPIENT\_COST:}} & Recipient cost of pull-style airdrop. All \airdroppees claim funds. \\
  \textbf{\textsf{\scriptsize BASE\_LINE:}} & Baseline for pull style airdrop, see Appendix~\ref{sec:simulation_setup}. \\
  \bottomrule
\end{tabularx}
\label{tbl:labels}
\end{table}

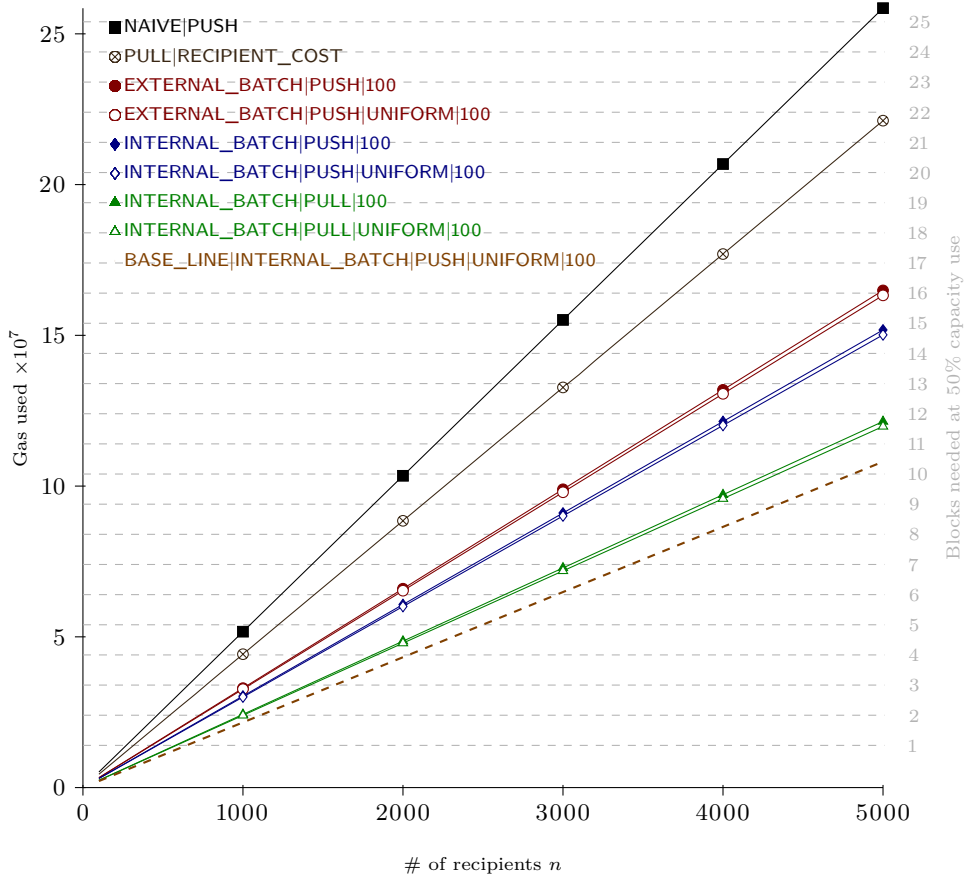
\begin{figure}[ht!]
  \begin{tikzpicture}[y=.004cm, x=.0021cm]
    \draw (0,0) -- coordinate (x axis mid) (5000,0);
    \draw (0,0) -- coordinate (y axis mid) (0,2585);  
        \foreach \x in {0,1000,...,5000}
          \draw (\x,1pt) -- (\x,-3pt)
            node[anchor=north] {\x};
        \foreach \y [evaluate=\y as \yy using int(ceil(\y*.01))] in {0,500,...,2585}
          \draw (1pt,\y) -- (-3pt,\y) 
            node[anchor=east] {\yy}; 
    \node[below=0.8cm] at (x axis mid) {\scriptsize \# of \airdroppees $n$};
    \node[rotate=90, above=.6cm, overlay] at (y axis mid) {\scriptsize Gas used $\times 10^7$};
    \node[rotate=90, above=-11.7cm, black!30] at (y axis mid) {\scriptsize Blocks needed at 50\% capacity use};

    \foreach \y in {1,2,...,25}
      \draw[dashed, black!30] (0, \y 39.98836)--++(5010,0) node [right, xshift=.2cm] {\scriptsize {\y}};

    \draw plot[mark=square*, mark phase=10, mark repeat=10]
      file {data/line_50_nodiscount_NAIVE_PUSH.csv};  

    \draw[red!50!black] plot[mark=*, mark phase=10, mark repeat=10]
      file {data/line_50_nodiscount_EXTERNAL_BATCH_PUSH_100.csv};  
    \draw[red!50!black] plot[mark=*, mark options={fill=white}, mark phase=10, mark repeat=10]
      file {data/line_50_nodiscount_EXTERNAL_BATCH_PUSH_ONE_AMOUNT_100.csv};  

    \draw[blue!50!black] plot[mark=diamond*, mark phase=10, mark repeat=10]
      file {data/line_50_nodiscount_INTERNAL_BATCH_PUSH_100.csv};  
    \draw[blue!50!black] plot[mark=diamond*, mark options={fill=white}, mark phase=10, mark repeat=10]
      file {data/line_50_nodiscount_INTERNAL_BATCH_PUSH_ONE_AMOUNT_100.csv};  

    \draw[green!50!black] plot[mark=triangle*, mark phase=10, mark repeat=10]
      file {data/line_50_nodiscount_INTERNAL_BATCH_PULL_100.csv};  
    \draw[green!50!black] plot[mark=triangle*, mark options={fill=white}, mark phase=10, mark repeat=10]
      file {data/line_50_nodiscount_INTERNAL_BATCH_PULL_ONE_AMOUNT_100.csv};

    \draw[brown!30!black] plot[mark=otimes*, mark phase=10, mark repeat=10, mark options={fill=white}]
      file {data/line_50_nodiscount_INTERNAL_BATCH_PULL_USER_COST.csv};  

    \draw[dashed, thick, orange!50!black] plot[]
      file {data/line_50_nodiscount_BASE_LINE_INTERNAL_BATCH_PUSH_ONE_AMOUNT_100.csv};

    \begin{scope}[shift={(200,1650)}] 

      \draw[yshift=9\baselineskip] (0,0) -- 
        plot[mark=square*] (0.25,0) -- (0.5,0)
        node[right,font=\sffamily]{\scriptsize NAIVE$$|$$PUSH};

      \draw[yshift=8\baselineskip, brown!30!black] (0,0) -- 
        plot[mark=otimes*,  mark options={fill=white}] (0.25,0) -- (0.5,0)
        node[right,font=\sffamily]{\scriptsize PULL$$|$$RECIPIENT\_COST};

      \draw[yshift=7\baselineskip, red!50!black] (0,0) -- 
        plot[mark=*] (0.25,0) -- (0.5,0)
        node[right,font=\sffamily]{\scriptsize EXTERNAL\_BATCH$$|$$PUSH$$|$$100};
      \draw[yshift=6\baselineskip, red!50!black] (0,0) -- 
        plot[mark=*, mark options={fill=white}] (0.25,0) -- (0.5,0)
        node[right,font=\sffamily]{\scriptsize EXTERNAL\_BATCH$$|$$PUSH$$|$$UNIFORM$$|$$100};

      \draw[yshift=5\baselineskip, blue!50!black] (0,0) -- 
        plot[mark=diamond*] (0.25,0) -- (0.5,0)
        node[right,font=\sffamily]{\scriptsize INTERNAL\_BATCH$$|$$PUSH$$|$$100};
      \draw[yshift=4\baselineskip, blue!50!black] (0,0) -- 
        plot[mark=diamond*, mark options={fill=white}] (0.25,0) -- (0.5,0)
        node[right,font=\sffamily]{\scriptsize INTERNAL\_BATCH$$|$$PUSH$$|$$UNIFORM$$|$$100};

      \draw[yshift=3\baselineskip, green!50!black] (0,0) -- 
        plot[mark=triangle*] (0.25,0) -- (0.5,0)
        node[right,font=\sffamily]{\scriptsize INTERNAL\_BATCH$$|$$PULL$$|$$100};
      \draw[yshift=2\baselineskip, green!50!black] (0,0) -- 
        plot[mark=triangle*, mark options={fill=white}] (0.25,0) -- (0.5,0)
        node[right,font=\sffamily]{\scriptsize INTERNAL\_BATCH$$|$$PULL$$|$$UNIFORM$$|$$100};

      \draw[yshift=\baselineskip, dashed, thick, orange!50!black] (0,0) -- 
        plot[] (0.25,0) -- (0.5,0)
        node[right,font=\sffamily]{\scriptsize BASE\_LINE$$|$$INTERNAL\_BATCH$$|$$PUSH$$|$$UNIFORM$$|$$100};

    \end{scope}
  \end{tikzpicture}
  \caption{Simulated cost for $n$ recipients using different strategies. The chart shows the subset of strategies whose transactions fit into 50\% of the block gas limit. Cost is not discounted, \ie we assume all \airdroppees are new to the \TOK, which makes the airdrop a bit more expensive.}
  \label{fig:methods_line}
\end{figure}

Figure~\ref{fig:methods_line} presents the gas cost as a function of the number of \airdroppees. We only show strategies viable when targeting a less than 50\% block fill grade. Observe that all strategies behave broadly linear. The fixed cost per batch is negligible. \approachlbl{NAIVE$|$PUSH} and \approachlbl[orange!50!black]{BASE\_LINE$|$INTERNAL\_BATCH$|$PUSH$|$UNIFORM$|$100} serve as upper and lower bounds and thus are benchmarks for the other strategies. The lower bound (baseline) simulates a push-style airdrop only considering communication and storage cost.\footnote{This rests on the assumption that other computation cost can be optimized. See Appendix~\ref{sec:simulation_setup} for more details.}

Although, \approachlbl[green!50!black]{INTERNAL\_BATCH$|$PULL$|$UNIFORM$|$100} appears to be the cheapest strategy, this is only true for the \airdropper. Recall that in pull-style airdrops \airdroppees have to send additional transactions in order to withdraw their tokens. If we sum up the costs of the \airdroppee (\approachlbl[brown!50!black]{PULL$|$RECIPIENT\_COST}) and \airdropper, the pull-style airdrop is by far the most expensive, \num{32}\% more costly than \approachlbl{NAIVE$|$PUSH}. 

The biggest improvement for both parties compared to the \approachlbl{NAIVE$|$PUSH} is archived by \approachlbl[blue!50!black]{INTERNAL\_BATCH$|$PUSH$|$UNIFORM$|$100}, saving roughly \num{42}\%. 
The baseline suggests that savings up to \num{58}\% are be possible.  
If we compare internal vs.\ external batching, the internal strategies are save around \num{8}\% compared to their externally batched counterparts. 
The uniform strategies only save about \num{1}\% compared to their counterparts. The savings might go up if larger amounts are transferred, which require more non-zero bytes to encode. However, the batching contract could support logarithmic scaling or batch-wide multipliers, which make our approximation with two non-zero bytes per amount realistic again. 

\begin{figure}[ht!]
  \centering
   \begin{tikzpicture}[>=stealth,x=.25cm,y=.015cm]

          \draw (0,0) -- coordinate (x axis mid) (44,0);
          \draw (0,0) -- coordinate (y axis mid) (0,525);  

          \node[rotate=90, above=.7cm, overlay] at (y axis mid) {\scriptsize Gas used $\times 10^7$};
          \draw (21,0) node [below, yshift=-10] {\scriptsize Strategies};


          \foreach \y [evaluate=\y as \yy using \y / 100] in  {0,50,...,520}
              \draw (2pt,\y)--++(-4pt,0) node [left] {\scriptsize \num[round-precision=1]{\yy}};


          \begin{scope}[line width=1.7mm]
              
            \draw [green!50!black] (1,0)--++(0,517.0488) node[above, xshift=-.1cm, rotate=45, black, anchor=west,font=\sffamily] {\tiny NAIVE$|$PUSH};

\draw [green!50!black] (2,0)--++(0,0);

\draw [green!50!black] (3,0)--++(0,442.4088) node[above, xshift=-.1cm, rotate=45, black, anchor=west,font=\sffamily] {\tiny PULL$|$RECIPIENT\_COST};

\draw [green!50!black] (4,0)--++(0,0);

\draw [blue!50!black] (5,0)--++(0,326.5184) node[above, xshift=-.1cm, rotate=45, black, anchor=west,font=\sffamily] {\tiny EXTERNAL\_BATCH$|$PUSH$|$UNIFORM$|$100};
\draw [red!50!black] (6,0)--++(0,325.3917) node[above, xshift=-.1cm, rotate=45, black, anchor=west,font=\sffamily] {\tiny 200};
\draw [black!30] (7,0)--++(0,325.16945) node[above, xshift=-.1cm, rotate=45, black, anchor=west,font=\sffamily] {\tiny 300};
\draw [black!30] (8,0)--++(0,324.94464) node[above, xshift=-.1cm, rotate=45, black, anchor=west,font=\sffamily] {\tiny 400};

\draw [green!50!black] (9,0)--++(0,0);

\draw [blue!50!black] (10,0)--++(0,329.7965) node[above, xshift=-.1cm, rotate=45, black, anchor=west,font=\sffamily] {\tiny EXTERNAL\_BATCH$|$PUSH$|$100};
\draw [red!50!black] (11,0)--++(0,328.6547) node[above, xshift=-.1cm, rotate=45, black, anchor=west,font=\sffamily] {\tiny 200};
\draw [black!30] (12,0)--++(0,328.43489) node[above, xshift=-.1cm, rotate=45, black, anchor=west,font=\sffamily] {\tiny 300};
\draw [black!30] (13,0)--++(0,328.20994) node[above, xshift=-.1cm, rotate=45, black, anchor=west,font=\sffamily] {\tiny 400};

\draw [green!50!black] (14,0)--++(0,0);

\draw [blue!50!black] (15,0)--++(0,300.309) node[above, xshift=-.1cm, rotate=45, black, anchor=west,font=\sffamily] {\tiny INTERNAL\_BATCH$|$PUSH$|$UNIFORM$|$100};
\draw [brown!50!black] (16,0)--++(0,299.17695) node[above, xshift=-.1cm, rotate=45, black, anchor=west,font=\sffamily] {\tiny 200};
\draw [black!30] (17,0)--++(0,298.95363) node[above, xshift=-.1cm, rotate=45, black, anchor=west,font=\sffamily] {\tiny 300};
\draw [black!30] (18,0)--++(0,298.72775) node[above, xshift=-.1cm, rotate=45, black, anchor=west,font=\sffamily] {\tiny 400};

\draw [green!50!black] (19,0)--++(0,0);

\draw [blue!50!black] (20,0)--++(0,303.5716) node[above, xshift=-.1cm, rotate=45, black, anchor=west,font=\sffamily] {\tiny INTERNAL\_BATCH$|$PUSH$|$100};
\draw [red!50!black] (21,0)--++(0,302.4322) node[above, xshift=-.1cm, rotate=45, black, anchor=west,font=\sffamily] {\tiny 200};
\draw [black!30] (22,0)--++(0,302.21284) node[above, xshift=-.1cm, rotate=45, black, anchor=west,font=\sffamily] {\tiny 300};
\draw [black!30] (23,0)--++(0,301.98836) node[above, xshift=-.1cm, rotate=45, black, anchor=west,font=\sffamily] {\tiny 400};

\draw [green!50!black] (24,0)--++(0,0);

\draw [blue!50!black] (25,0)--++(0,239.5702) node[above, xshift=-.1cm, rotate=45, black, anchor=west,font=\sffamily] {\tiny INTERNAL\_BATCH$|$PULL$|$UNIFORM$|$100};
\draw [brown!50!black] (26,0)--++(0,238.44255) node[above, xshift=-.1cm, rotate=45, black, anchor=west,font=\sffamily] {\tiny 200};
\draw [red!50!black] (27,0)--++(0,238.22011) node[above, xshift=-.1cm, rotate=45, black, anchor=west,font=\sffamily] {\tiny 300};
\draw [black!30] (28,0)--++(0,237.99511) node[above, xshift=-.1cm, rotate=45, black, anchor=west,font=\sffamily] {\tiny 400};

\draw [green!50!black] (29,0)--++(0,0);

\draw [blue!50!black] (30,0)--++(0,242.8482) node[above, xshift=-.1cm, rotate=45, black, anchor=west,font=\sffamily] {\tiny INTERNAL\_BATCH$|$PULL$|$100};
\draw [brown!50!black] (31,0)--++(0,241.7055) node[above, xshift=-.1cm, rotate=45, black, anchor=west,font=\sffamily] {\tiny 200};
\draw [red!50!black] (32,0)--++(0,241.48548) node[above, xshift=-.1cm, rotate=45, black, anchor=west,font=\sffamily] {\tiny 300};
\draw [black!30] (33,0)--++(0,241.26034) node[above, xshift=-.1cm, rotate=45, black, anchor=west,font=\sffamily] {\tiny 400};

\draw [green!50!black] (34,0)--++(0,0);

\draw [blue!50!black] (35,0)--++(0,216.18256) node[above, xshift=-.1cm, rotate=45, black, anchor=west,font=\sffamily] {\tiny BASE\_LINE$|$INTERNAL\_BATCH$|$PUSH$|$UNIFORM$|$100};
\draw [brown!50!black] (36,0)--++(0,215.13256) node[above, xshift=-.1cm, rotate=45, black, anchor=west,font=\sffamily] {\tiny 200};
\draw [red!50!black] (37,0)--++(0,214.92256) node[above, xshift=-.1cm, rotate=45, black, anchor=west,font=\sffamily] {\tiny 300};
\draw [black!30] (38,0)--++(0,214.71256) node[above, xshift=-.1cm, rotate=45, black, anchor=west,font=\sffamily] {\tiny 400};
\draw [black!30] (39,0)--++(0,214.50256) node[above, xshift=-.1cm, rotate=45, black, anchor=west,font=\sffamily] {\tiny 500};
\draw [black!30] (40,0)--++(0,214.50256) node[above, xshift=-.1cm, rotate=45, black, anchor=west,font=\sffamily] {\tiny 600};
\draw [black!30] (41,0)--++(0,214.50256) node[above, xshift=-.1cm, rotate=45, black, anchor=west,font=\sffamily] {\tiny 700};
\draw [black!30] (42,0)--++(0,214.50256) node[above, xshift=-.1cm, rotate=45, black, anchor=west,font=\sffamily] {\tiny 800};
          \end{scope}






        \begin{scope}[shift={(30,500)}] 

          \draw[yshift=\baselineskip, black!30, font=\sffamily] (0,0) -- 
            plot[mark=square*] (0.25,0) -- (0.5,0)
            node[right]{\scriptsize Not feasible};

          \draw[yshift=2\baselineskip, red!50!black, font=\sffamily] (0,0) -- 
            plot[mark=square*] (0.25,0) -- (0.5,0)
            node[right]{\scriptsize Feasible within 100\% gas limit};

          \draw[yshift=3\baselineskip, brown!50!black, font=\sffamily] (0,0) -- 
            plot[mark=square*] (0.25,0) -- (0.5,0)
            node[right]{\scriptsize Feasible within 75\% gas limit};

          \draw[yshift=4\baselineskip, blue!50!black, font=\sffamily] (0,0) -- 
            plot[mark=square*] (0.25,0) -- (0.5,0)
            node[right]{\scriptsize Feasible within 50\% gas limit};

          \draw[yshift=5\baselineskip, orange!50!black, font=\sffamily] (0,0) -- 
            plot[mark=square*] (0.25,0) -- (0.5,0)
            node[right]{\scriptsize Feasible within 25\% gas limit};

          \draw[yshift=6\baselineskip, green!50!black, font=\sffamily] (0,0) -- 
            plot[mark=square*] (0.25,0) -- (0.5,0)
            node[right]{\scriptsize Feasible within 10\% gas limit};

        \end{scope}
        
    \end{tikzpicture}
  \caption{Cost and feasibility of different strategies with \num{1000} recipients with different cutoffs on the block gas limit.}
  \label{fig:methods_bar}
\end{figure}
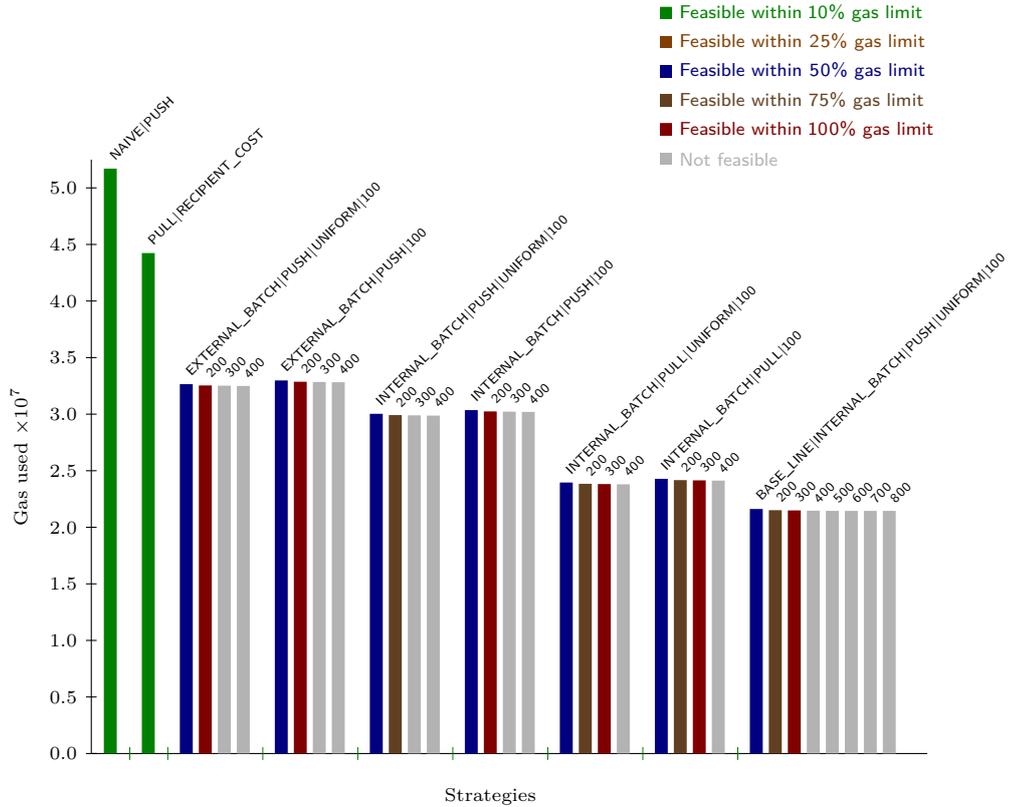

Figure~\ref{fig:methods_bar} shows a single simulation run for \num{1000} \airdroppees. 
Given the approximate linearity the number of \airdroppees, this view is sufficient to compare the strategies. This time we present all strategies.
We color code the minimal block fill grade in which each strategy gets feasible.
The first thing to observe is that batches of more than 300~\airdroppees are not feasible with current block gas limits\footnote{The block gas limit used as cutoff can be found in Table~\ref{table:constants}. Appendix~\ref{sec:simulation_setup}.}. 
Only the pull approaches and the baseline can manage a batch size of up to \num{300}.
Note that \approachlbl{NAIVE$|$PUSH} as well as the \approachlbl{RECIPIENT\_COST} are feasible even with a threshold of \num{10}\% block fill grade, since no batching is applied.  

We did not simulate the pooled payment strategy (see \ref{sec:pooledpay}). Since the \airdropper cost is constant by only storing the Merkle root, this would make it the by far cheapest option for \airdroppers. The \airdroppees have to withdraw the tokens in a very similar manner to the \approachlbl[brown!50!black]{PULL$|$RECIPIENT\_COST} strategy. In addition, the \airdroppee has to pay for the verification of the Merkle proof and the storage of the withdrawal record.

\section{Discussion}
\label{sec:discussion}

The above results show differences in gas consumption. More relevant units of operational cost for the \airdropper are ETH, if the \airdropper is already invested in \ET, and fiat (USD). To put our results into perspective, we estimate the cost savings per \num{1000} \airdroppees in USD. This entails two conversions with variable rates: from gas to ETH and from ETH to USD. The first conversion is governed by the fee market mechanism and the miner's transaction inclusion strategy. Since transactions compete for inclusion in the chain, the gas price (in ETH) depends on the network load. The second conversion rate is found on the cryptocurrency exchanges in the ecosystem. The price depends on demand and supply of cryptocurrency, which supposedly follow investors' economic expectations.

\begin{figure}[ht!]
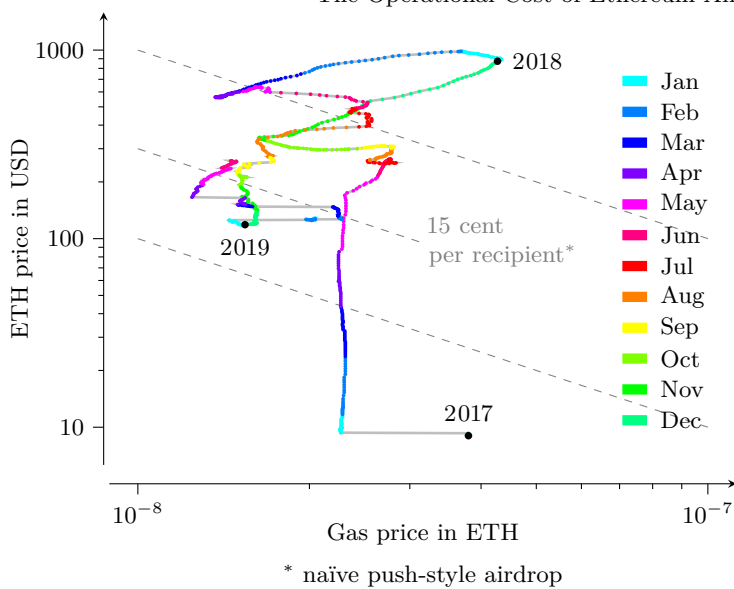

\begin{center}
\begin{tikzpicture}[>=stealth,x=7.5cm,y=2.5cm]


\definecolor{col7}{RGB}{255,0,0}
 \definecolor{col8}{RGB}{255,128,0}
 \definecolor{col9}{RGB}{255,255,0}
 \definecolor{col10}{RGB}{128,255,0}
 \definecolor{col11}{RGB}{0,255,0}
 \definecolor{col12}{RGB}{0,255,128}
 \definecolor{col1}{RGB}{0,255,255}
 \definecolor{col2}{RGB}{0,128,255}
 \definecolor{col3}{RGB}{0,0,255}
 \definecolor{col4}{RGB}{128,0,255}
 \definecolor{col5}{RGB}{255,0,255}
 \definecolor{col6}{RGB}{255,0,128}
 
	
	\draw [dashed,gray] (-8,3)--++(1,-1); 
	\draw [dashed,gray] (-8,2)--++(1,-1); 
	\draw [dashed,gray] (-8,2.477)-- node [right,pos=.99] {\parbox{2cm}{15 cent\\per recipient$^*$}} ++(.5,-.5); 

	
	\draw [->] (-8.05,.7) coordinate (xorig) -- 
			node [below=3ex] {Gas price in ETH}
			(-6.95,.7);
	\draw (-8,0|-xorig)--++(0,-4pt) node [below] {\small $10^{-8}$};
	\draw (-7,0|-xorig)--++(0,-4pt) node [below] {\small $10^{-7}$};
	
	\foreach \x in {0.301,0.4771,0.6021,0.699,0.7782,0.8451,0.9031,0.9542}
		\draw (-8,0|-xorig)++(\x,0)--++(0,-2pt);
		
	
	\begin{scope}[overlay]
	\draw [->] (-8.06,.8) coordinate (yorig) -- 
		node [rotate=90,above=6ex] {ETH price in USD}
		(-8.06,3.2);
	\draw (0,1-|yorig)--++(-4pt,0) node [left] {\small $10$};
	\draw (0,2-|yorig)--++(-4pt,0) node [left] {\small $100$};
	\draw (0,3-|yorig)--++(-4pt,0) node [left] {\small $1000$};

	\foreach \x in {0.301,0.4771,0.6021,0.699,0.7782,0.8451,0.9031,0.9542}
	{
		\draw (0,1-|yorig)++(0,\x)--++(-2pt,0);
		\draw (0,2-|yorig)++(0,\x)--++(-2pt,0);
	}
	\end{scope}


	\begin{scope}[line width=1pt,line cap=round,black!25]
	\input{figures/gaspath}
	\input{figures/gascoldots}
	\end{scope}
	
	
	\draw (M2017-01) node [fill,circle,inner sep=1pt] {~} node [above=2pt] {\small 2017};
	\draw (M2018-01) node [fill,circle,inner sep=1pt] {~} node [right=2pt] {\small 2018};
	\draw (M2019-01) node [fill,circle,inner sep=1pt] {~} node [below=2pt] {\small 2019};
	
	
	\foreach \c/\m in {1/Jan,2/Feb,3/Mar,4/Apr,5/May,6/Jun,7/Jul,8/Aug,9/Sep,10/Oct,11/Nov,12/Dec}
		\draw (-7.15,3)++(0,-\c ex)++(0,-\c ex)++(0,-\c ex) [col\c,line width=1ex] --++(1em,0) node [right,black] {\small\m};

\end{tikzpicture}
\\$^*$ na\"ive push-style airdrop
\caption{Dynamics of the price of \ET resource use, broken down into components. The plot shows 60-day moving averages of daily prices reported by Etherscan.io. Dashed lines connect levels of equal gas price in USD.}
\label{fig:gasethusd}
\end{center}
\end{figure}

We do not aim to explain price formation in this work (although airdrops may affect prices in the short run), but take an empirical approach. Figure~\ref{fig:gasethusd} shows the co-movement of both prices from January 2017 to June 2019 on a log-log scale. We calculate 60-day moving averages and represent each center day as dot, color-coded by the calendar month. The dashed lines connect levels of equal gas price in USD. Observe that both prices follow different dynamics, hence are not strongly coupled by a single market mechanism. While the ETH/USD exchange rate varies over two orders of magnitude in the sample period, the gas price in ETH remains in a much narrower band. However, it exhibits more sudden jumps, which relate to extreme values (\eg due to congestion) that enter or leave the moving window. One can also speculate if the introduction of gas tokens in spring 2018 has narrowed the band of gas price movement due to the counter-cyclical behavior of gas token investors.

To get an idea of airdrop costs in USD, the dashed line marked with \num{15} cents (of USD) per \airdroppee seems a good rule of thumb. This price level was applicable for a na\"ive push-style airdrop in May 2017, October 2018, and in March and May 2019. More efficient strategies have costed around \num{7.5} cents per \airdroppee. Given the variability of both prices (some of which is hidden by the moving average), it seems that the right timing is at least as important as the strategy.

To continue the example from above, the OmiseGO airdrop distributed tokens to \num{450000} \airdroppees, using an externally batched push-style approach. By applying the gas-to-USD conversion rate indicated in Figure~\ref{fig:gasethusd}, we can estimate the cost at roughly \num{44523} USD. Airdrops of this size occupy 50\% of the available capacity in \num{1440} blocks, taking at least \num{6} hours to complete. As a consequence, early \airdroppees have an advantage when selling tokens on an exchange immediately after receipt. This suggests that \TOKS should support time locks in order to enable large and \emph{fair} airdrops.

\section{Related Work}
\label{sec:related_work}

Our work connects to prior works on the systematic analysis of token systems on the \EP, gas efficiency, and one seminal publication on airdrops.

\textbf{Token Systems and ICOs:}
Howell~\EA\cite{howell2018initial} study the success factors of 440 ICOs on \ET based on propriety transaction data, presumably acquired from exchanges and other intermediaries, and manual labeling.
Their main interest is in the relationship between issuer characteristics and indicators of success.
The regression analyses find highly significant positive effects on liquidity and volume of the token for  independent variables measuring the existence of a white paper, the availability of code on Github, the support by venture capitalists, the entrepreneurs' experience, the acceptance of Bitcoin, and the organization of a pre-sale. No significant effect is found for airdrops.

Friedhelm~\EA\cite{victormeasuring} study \ET \TOKS from a network perspective.
They find that the degree distribution of the token network transfers does not follow a power law, but is dominated by a few hubs. In particular recipients of initial tokens mainly trade with these hubs. Some tokens systems seem to be very illiquid. Airdrops are not considered.

\textbf{Gas Usage:}
Chen~\EA\cite{chen2017adaptive} identify underpriced instructions (even after the 2016 gas price adjustment) and propose an adaptive pricing scheme. Their main interest is to raise economic barriers against congestion, which in the worst case enables denial of service attacks on the systemic level. 

In a different work, Chen~\EA\cite{chen2017under} use pattern matching to identify code sequences that can be further optimized for gas use in smart contracts deployed until 2016. Naegele and Schett~\cite{nageleblockchain} pursue a similar goal with the help of SMT solvers. Both source report ample room for improvement. 
While the referenced works optimize on the instruction level, the optimizations studied in this paper primarily seek to minimize communication overhead.

\textbf{Airdrops:}
Airdrops are a rather new topic. We are aware of one academic paper only.
Harrigan~\EA\cite{harrigan2018airdrops} raises awareness for privacy implications of airdrops when identifiers of one chain (\emph{source chain}) are reused to distribute coins on another chain (\emph{target chain}). Sharing identifiers between chains in general gives additional clues for address clustering.

To the best of our knowledge, we are the first to compare the gas costs of technical solutions for airdrops of tokens on the \EP.

\section{Conclusion and Outlook}
\label{sec:conclusion}

This work compared the efficiency of bulk transfer approaches on the \EP, a general problem that became particularly relevant with the uptake of token airdrops.
It turns out that many of the approaches we systematized and reviewed are workarounds for architectural short-comings of the \EP or the popular \ERC standard for fungible tokens.
The cost efficiency of the approaches differ roughly by a factor of two. Moreover, the most cost-efficient solutions for the \airdropper impose significant cost on the \airdroppee, which might thwart the very intention of an airdrop as marketing tool. We release our simulation framework and the model contracts for reproducibility, as testbed for actual airdrops, and as benchmarking suite for new solutions.

The choice of approach is constrained by properties of the \TOK. This mainly relates to the penalty of repeatedly calling a method from \emph{another} contract, which appears disproportional to the computational effort of the node. While a remote call is indeed expensive at first use, every repeated call is sped up through caching. \ET's gas price schedule seems to unfairly discriminate against bulk operations, an issue that designers of price schedules for future blockchain platforms should fix. On \ET as it stands, token issuers are best advised to reflect about airdrops before deployment of the \TOK contract. Planning ahead is vital in an environment where code cannot be amended easily. 



While framed and motivated for the application of airdrops, our analysis generalizes to any kind of bulk operation on lists of incompressible items. Future designs of blockchain platforms should consider mitigating most of the issues discussed here by supporting a global index for constants with high entropy. 
Some cryptographic material, in particular public keys and commitments, must be stored on-chain in order to enable authorization of actions by the knowledge of secrets. But if they do not double as references, as in \ET, then every value has to be stored (and paid for) only once. Furthermore, if all contracts have access to all public information on the blockchain, sets can be reused and storage space saved. In short, the community needs a DRY (don't repeat yourself) principle for data on the blockchain.

\section*{Acknowledgments}
This work has received funding from the European Union's Horizon 2020 research and innovation programme under grant agreement No.\ 740558.

\bibliographystyle{splncs03}
\bibliography{paper}

\begin{thebibliography}{10}
\providecommand{\url}[1]{\texttt{#1}}
\providecommand{\urlprefix}{URL }

\bibitem{ercstd}
{EIP 20: ERC-20 Token Standard}. \\\url{https://eips.ethereum.org/EIPS/eip-20}
  (2015), {[Online; accessed 18 Jun 2019]}

\bibitem{OmiseGOAD}
{OmiseGO tokens airdrop}. \\\url{https://github.com/omisego/airdrop} (2017),
  [Online; accessed 18 Jun 2019]

\bibitem{AirdropGasCrisis}
{Ethereum’s Growing Gas Crisis (And What’s Being Done to Stop It)}.
  \\\url{https://www.coindesk.com/ethereums-growing-gas-crisis-and-whats-being-done-to-stop-it}
  (2018), [Online; accessed 18 Jun 2019]

\bibitem{MerkleMine}
{MerkleMine Specification}.
  \\\url{https://github.com/livepeer/merkle-mine/blob/master/SPEC.md} (2018),
  [Online; accessed 18 Jun 2019]

\bibitem{PooledPayments}
{Pooled Payments (scaling solution for one-to-many transactions)}.
  \\\url{https://ethresear.ch/t/pooled-payments-scaling-solution-for-one-to-many-transactions/590}
  (2018), [Online; accessed 18 Jun 2019]

\bibitem{Boehme2013-IAB}
B{\"o}hme, R.: Internet protocol adoption: {L}earning from {B}itcoin. IAB
  Workshop on Internet Technology Adoption and Transition (ITAT), Cambridge
  (2013)

\bibitem{Briscoe2006}
Briscoe, B., Odlyzko, A., Tilly, B.: Metcalfe's law is wrong. IEEE Spectrum
  43(7),  34--39 (2006)

\bibitem{chen2017under}
Chen, T., Li, X., Luo, X., Zhang, X.: Under-optimized smart contracts devour
  your money. In: 2017 IEEE 24th International Conference on Software Analysis,
  Evolution and Reengineering (SANER). pp. 442--446. IEEE (2017)

\bibitem{chen2017adaptive}
Chen, T., Li, X., Wang, Y., Chen, J., Li, Z., Luo, X., Au, M.H., Zhang, X.: An
  adaptive gas cost mechanism for ethereum to defend against under-priced dos
  attacks. In: International Conference on Information Security Practice and
  Experience. pp. 3--24. Springer (2017)

\bibitem{daian2019flash}
Daian, P., Goldfeder, S., Kell, T., Li, Y., Zhao, X., Bentov, I., Breidenbach,
  L., Juels, A.: Flash boys 2.0: Frontrunning, transaction reordering, and
  consensus instability in decentralized exchanges. arXiv preprint
  arXiv:1904.05234  (2019)

\bibitem{frowis2018detecting}
Fr{\"o}wis, M., Fuchs, A., B{\"o}hme, R.: Detecting token systems on ethereum.
  arXiv preprint arXiv:1811.11645  (2018)

\bibitem{harrigan2018airdrops}
Harrigan, M., Shi, L., Illum, J.: Airdrops and privacy: A case study in
  cross-blockchain analysis. In: 2018 IEEE International Conference on Data
  Mining Workshops (ICDMW). pp. 63--70. IEEE (2018)

\bibitem{Hill2006}
Hill, S., Provost, F., Volinsky, C.: Network-based marketing: {I}dentifying
  likely adopters via consumer networks. Statistical Science  21(2),  256--276
  (2006)

\bibitem{howell2018initial}
Howell, S.T., Niessner, M., Yermack, D.: Initial coin offerings: Financing
  growth with cryptocurrency token sales. Tech. rep., National Bureau of
  Economic Research (2018)

\bibitem{Katz1994}
Katz, M.L., Shapiro, C.: Systems competition and network effects. Journal of
  Economic Perspectives  8(2),  93--115 (1994)

\bibitem{merkle1987digital}
Merkle, R.C.: A digital signature based on a conventional encryption function.
  In: Conference on the theory and application of cryptographic techniques. pp.
  369--378. Springer (1987)

\bibitem{nageleblockchain}
Nagele, J., Schett, M.A.: Blockchain superoptimizer.
  \\\url{http://www.maria-a-schett.net/talks/2019_04-MaS_imdea.pdf} (2018),
  [Online; accessed 24 Jun 2019]

\bibitem{OMISEDetails}
nharrison: Airdrop update 2.
  \\\url{https://steemit.com/ethereum/@nharrison/what-you-need-to-know-about-the-omisego-airdrop}
  (2017), [Online; accessed 25 Jun 2019]

\bibitem{victormeasuring}
Victor, F., L{\"u}ders, B.K.: Measuring ethereum-based erc20 token networks.
  In: International Conference on Financial Cryptography and Data Security
  (2018)

\bibitem{approvalVulnerable}
Vladimirov, M., Khovratovich, D.: Erc20 api: An attack vector on
  approve/transferfrom methods.
  \\\url{https://docs.google.com/document/d/1YLPtQxZu1UAvO9cZ1O2RPXBbT0mooh4DYKjA_jp-RLM},
  [Online; accessed 25 Jun 2019]

\end{thebibliography}

\appendix

\clearpage

\section{Simulation Environment}
\label{sec:simulation_setup}

For the simulation, we use \texttt{ganache-cli},\footnote{Version 6.4.4; \url{https://github.com/trufflesuite/ganache-cli}} an implementation of an \ET node specifically designed for development and testing of smart contracts.
The Ethereum virtual machine of \texttt{ganache} builds on \texttt{ethereumjs}. 
We implement the simulation in Javascript and run it on \texttt{nodejs} version \texttt{v8.15.0}. The \texttt{web3} library serves as interface to the node.

\begin{figure}[ht!]
  \begin{tikzpicture}[y=.004cm, x=.0021cm]
    \draw (0,0) -- coordinate (x axis mid) (5000,0);
    \draw (0,0) -- coordinate (y axis mid) (0,2585);  
        \foreach \x in {0,1000,...,5000}
          \draw (\x,1pt) -- (\x,-3pt)
            node[anchor=north] {\x};
        \foreach \y [evaluate=\y as \yy using int(ceil(\y*.01))] in {0,500,...,2585}
          \draw (1pt,\y) -- (-3pt,\y) 
            node[anchor=east] {\yy}; 
    \node[below=0.8cm] at (x axis mid) {\scriptsize Nr of \airdroppees};
    \node[rotate=90, above=.6cm, overlay] at (y axis mid) {\scriptsize Gas used $\times 10^7$};
    \node[rotate=90, above=-11.7cm, black!30] at (y axis mid) {\scriptsize Blocks needed with 50\% fill};

    \foreach \y in {1,2,...,25}
      \draw[dashed, black!30] (0, \y 39.98836)--++(5010,0) node [right, xshift=.2cm] {\scriptsize {\y}};

    \draw plot[mark=square*, mark phase=10, mark repeat=10]
      file {data/line_50_discounted_value_NAIVE_PUSH.csv};  

    \draw[red!50!black] plot[mark=*, mark phase=10, mark repeat=10]
      file {data/line_50_discounted_value_EXTERNAL_BATCH_PUSH_100.csv};  
    \draw[red!50!black] plot[mark=*, mark options={fill=white}, mark phase=10, mark repeat=10]
      file {data/line_50_discounted_value_EXTERNAL_BATCH_PUSH_ONE_AMOUNT_100.csv};  

    \draw[blue!50!black] plot[mark=diamond*, mark phase=10, mark repeat=10]
      file {data/line_50_discounted_value_INTERNAL_BATCH_PUSH_100.csv};  
    \draw[blue!50!black] plot[mark=diamond*, mark options={fill=white}, mark phase=10, mark repeat=10]
      file {data/line_50_discounted_value_INTERNAL_BATCH_PUSH_ONE_AMOUNT_100.csv};  


    \draw[brown!30!black] plot[mark=otimes*, mark phase=10, mark repeat=10, mark options={fill=white}]
      file {data/line_50_discounted_value_INTERNAL_BATCH_PULL_USER_COST.csv};  

    \draw[dashed, thick, orange!50!black] plot[]
      file {data/line_50_discounted_value_BASE_LINE_INTERNAL_BATCH_PUSH_ONE_AMOUNT_100.csv};

    \begin{scope}[shift={(200,1800)}] 

      \draw[yshift=7\baselineskip] (0,0) -- 
        plot[mark=square*] (0.25,0) -- (0.5,0)
        node[right,font=\sffamily]{\scriptsize NAIVE$$|$$PUSH};

      \draw[yshift=6\baselineskip, brown!30!black] (0,0) -- 
        plot[mark=otimes*,  mark options={fill=white}] (0.25,0) -- (0.5,0)
        node[right,font=\sffamily]{\scriptsize PULL$$|$$RECIPIENT\_COST};

      \draw[yshift=5\baselineskip, red!50!black] (0,0) -- 
        plot[mark=*] (0.25,0) -- (0.5,0)
        node[right,font=\sffamily]{\scriptsize EXTERNAL\_BATCH$$|$$PUSH$$|$$100};
      \draw[yshift=4\baselineskip, red!50!black] (0,0) -- 
        plot[mark=*, mark options={fill=white}] (0.25,0) -- (0.5,0)
        node[right,font=\sffamily]{\scriptsize EXTERNAL\_BATCH$$|$$PUSH$$|$$UNIFORM$$|$$100};

      \draw[yshift=3\baselineskip, blue!50!black] (0,0) -- 
        plot[mark=diamond*] (0.25,0) -- (0.5,0)
        node[right,font=\sffamily]{\scriptsize INTERNAL\_BATCH$$|$$PUSH$$|$$100};
      \draw[yshift=2\baselineskip, blue!50!black] (0,0) -- 
        plot[mark=diamond*, mark options={fill=white}] (0.25,0) -- (0.5,0)
        node[right,font=\sffamily]{\scriptsize INTERNAL\_BATCH$$|$$PUSH$$|$$UNIFORM$$|$$100};


      \draw[yshift=\baselineskip, dashed, thick, orange!50!black] (0,0) -- 
        plot[] (0.25,0) -- (0.5,0)
        node[right,font=\sffamily]{\scriptsize BASE\_LINE$$|$$INTERNAL\_BATCH$$|$$PUSH$$|$$UNIFORM$$|$$100};

    \end{scope}
  \end{tikzpicture}
  \caption{Discounted version of Figure~\ref{fig:methods_line}. Discounted pull-style airdrop has been removed from the plot since it would be vulnerable to double spending, see Section \ref{sec:pull}.}
  \label{fig:methods_line_discounted}
\end{figure}
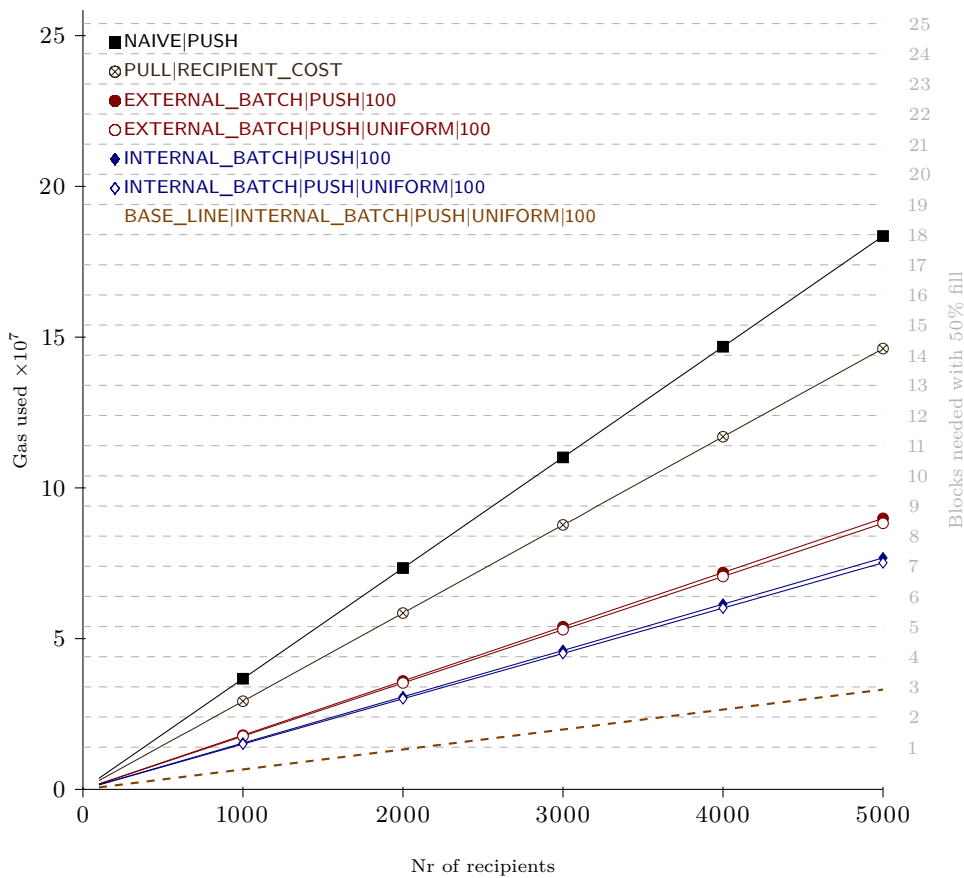

Every simulation run deploys a fresh versions of the simulated contracts and generates new accounts for all \airdroppees as well as the \airdropper. Consequently, each transfer requires a new storage slot for the \airdroppee. This costs \num{20000} gas, in contrast to \num{5000} if an additional token is transferred to a \airdroppee who already owns tokens of that type. To distinguish these two cases, we calculate a \textbf{discounted} scenario by subtracting the gas cost difference of \num{15000} per \airdroppee. We report the discounted results in Figure~\ref{fig:methods_line_discounted} for completeness. This scenario to some extent contradicts the purpose of an airdrop, which is to distribute tokens to new owners.

In all scenarios we assume a two-byte number of tokens to be distributed to all \airdroppees. This is relevant because the number of zero bytes in the transaction input influences the gas cost. If the number of tokens per \airdroppee requires more than two bytes, the cost gap between uniform and non-uniform distribution grows. External and internal batch transfer functions can reduce the cost of non-uniform distributions of highly divisible tokens by implementing a batch-wide amount multiplier. We have not measured this option. 

Table~\ref{table:constants} shows the constants used in our simulation and analysis along with their source.

\begin{table}
\centering
  \caption{Constants used in the simulation and analysis.}
\label{table:constants}
  \begin{tabular}{ lrp{1cm}p{4.5cm} } 
	\toprule
	\textbf{Name} & \multicolumn{1}{l}{\textbf{Value}} & \textbf{Unit} & \textbf{Source}\\
\midrule
  \textbf{Low gas price} & \num{0.58} & Wei & \url{https://ethgasstation.info/}, [Online; accessed 21 Jun 2019] \\
  \textbf{Median gas price} & \num{10.5}\phantom{0} & Wei & \url{https://ethgasstation.info/}, [Online; accessed 21 Jun 2019] \\
  \textbf{High gas price} & \num{235}\phantom{.00} & Wei & \url{https://ethgasstation.info/}, [Online; accessed 21 Jun 2019] \\
  \textbf{Block gas limit} & \num{7997671}\phantom{.00} & gas & Mean over all (main chain) blocks in 2018\\ 
\bottomrule
  \end{tabular}
\end{table}

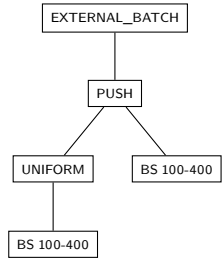
\begin{figure}[ht]
\centering
  \begin{tikzpicture}[sibling distance=20em, level distance=1cm,
    every node/.style = {shape=rectangle,
      draw, align=center,
      top color=white, bottom color=white},font=\tiny\sffamily, level 1/.style={sibling distance=15em}, level 2/.style={sibling distance=5em}]

       \node {EXTERNAL\_BATCH} 
            child { node {PUSH} 
                  child { node {UNIFORM}
                        child { node {BS 100-400}}
                  }
                  child { node {BS 100-400}}
            };
  \end{tikzpicture}
  \caption{Externally batched airdrop strategies (8) that were run in the simulation. BS stands for \emph{Batch Size}. Batch size steps are always \num{100}.}
  \label{fig:simspace1}
\end{figure}

Figures \ref{fig:simspace1},  \ref{fig:simspace2}, and \ref{fig:simspace3} illustrate our approach to systematically generate the \num{35} simulation scenarios. Every path from the root to a leaf node represents one of the strategies evaluated. Whenever a node contains a range of batch sizes (BS), the actual size is varied in steps of 100 in separate scenarios. Equation~\ref{eq:1} documents how the \textsf{\scriptsize BASE\_LINE|INTERNAL\_BATCH|PUSH|UNIFORM} strategies are computed $G_{baseline}(1000, 100)$ calculates the baseline cost for 1000 \airdroppees and a batch size of \num{100}.

\begin{mycapequ}[!ht] 
  \begin{equation}
    \begin{split}
        \operatorname{G_{tx}} &= 21000 \\
        \operatorname{G_{sstorenew}} &= 20000 \\
        \operatorname{G_{zeroinput}} &= 4 \\
        \operatorname{G_{nonzeroinput}}  &= 68 \\
        \operatorname{n_{tx}}(n, \mathit{bs}) &= \ceil*{\frac{n}{\mathit{bs}}} \\
        \operatorname{G_{inputword}}(b_{set}) &= \operatorname{G_{nonzeroinput}} \cdot\, b_{set} + \operatorname{G_{zeroinput}} \cdot\, (32 - b_{set})\\
        \operatorname{G_{inputuniform}}(n) &= (n \cdot  \operatorname{G_{inputword}}(20)) +  \operatorname{G_{inputword}}(2)  \\
        \operatorname{G_{sstores}}(n) &= n \cdot \operatorname{G_{sstorenew}} \\ 
        \operatorname{G_{txs}}(n, \mathit{bs}) &=  \operatorname{n_{tx}}(n, bs) \cdot \operatorname{G_{tx}}\\
        \operatorname{G_{baseline}}(n, bs) &= \operatorname{G_{txs}}(n, bs) + \operatorname{G_{sstore}}(n) + \operatorname{G_{inputuniform}}(n)
    \end{split}
    \label{eq:1}
  \end{equation}
  \caption{Equation 1: Calculation of baseline strategy. Only transaction overhead, transaction input and storage writes are considered. Input sizes are hard coded to 20 bytes per address and 2 bytes for the amount.}
\end{mycapequ}

\begin{figure}[ht]
\centering
  \begin{tikzpicture}[sibling distance=10em, level distance=1cm,
    every node/.style = {shape=rectangle,
      draw, align=center,
      top color=white, bottom color=white},font=\tiny\sffamily, level 1/.style={sibling distance=15em}, level 2/.style={sibling distance=6em}]
      \node {INTERNAL\_BATCH} 
            child { node {PUSH} 
                  child { node {UNIFORM}
                        child { node {BS 100-400}}
                  }
                  child { node {BS 100-400}}
            }
            child { node {PULL} 
                  child { node {UNIFORM}
                        child { node {BS 100-400}}
                        child { node {UNIFORM, BS 1}}
                  }
                  child { node {BS 100-400}}
                  child { node {RECIPIENT\_COST, BS 1}}
            };
  \end{tikzpicture}
  \caption{Internally batched airdrop strategies (18) that were run in the simulation. BS stands for \emph{Batch Size}. Batch size steps are always \num{100}.}
  \label{fig:simspace2}
\end{figure}
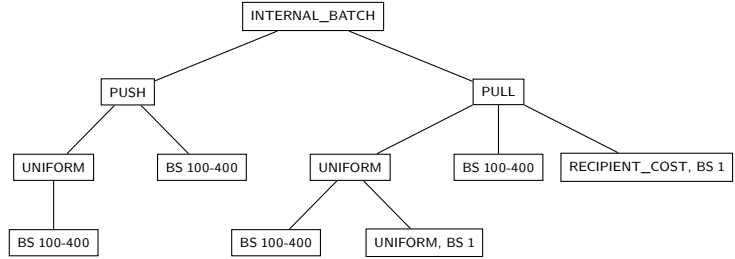

\begin{figure}[ht]
\centering
  \begin{tikzpicture}[sibling distance=10em, level distance=1cm,
    every node/.style = {shape=rectangle,
      draw, align=center,
      top color=white, bottom color=white},font=\tiny\sffamily, level 1/.style={sibling distance=6em}, level 2/.style={sibling distance=9em}]

    \node {\textcolor{red!50!black}{Upper}- and \textcolor{green!50!black}{lower}-bounds}
      child[dashed] { node [red!50!black, solid] {NAIVE} 
              child[solid] { node [red!50!black] {PUSH}
                    child { node [red!50!black] {BS 1} }
              }
            }
      child[dashed] { node [green!50!black, solid] {INTERNAL\_BATCH} 
            child[solid] { node [green!50!black] {PUSH} 
                  child { node [green!50!black] {UNIFORM}
                        child { node [green!50!black] {BS 100-800} child { node [green!50!black] {BASE\_LINE}}}
                  }
            }
      };
  \end{tikzpicture}
  \caption{Airdrop strategies (9) that serve as \textcolor{red!50!black}{upper}- and \textcolor{green!50!black}{lower}-bound in the analysis. BS stands for \emph{Batch Size}. Batch size steps are always \num{100}.}
  \label{fig:simspace3}
\end{figure}
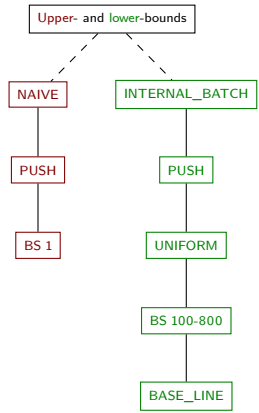

The complete simulation code along with analysis scripts and visualization will be released on Github.\footnote{\url{https://github.com/soad003/TheOperationalCostOfEthereumAirdrops}}

\section{Contracts Used in the Simulation}
\label{sec:contracts}

For our simulation, we use a slightly modified version of the popular OpenZeppelin implementation of an \ERC token\footnote{\url{https://github.com/OpenZeppelin/openzeppelin-solidity/blob/9b3710465583284b8c4c5d2245749246bb2e0094/contracts/token/ERC20/ERC20.sol}, commit: 9b3710465583284b8c4c5d2245749246bb2e0094}. 
We add internal batching to the \ERC token by copying the functions \soltext{airdrop}, \soltext{airdropDynamic}, \soltext{airdropApprove}, and \soltext{airdropApproveDynamic} from the external batching (\texttt{Airdropper.sol}) contract into the \ERC token contract. The external batching contract was inspired by a real batching contract.\footnote{\url{https://github.com/iosiro/airdropper/blob/master/contracts/Airdropper.sol}, commit: 3667ec866a5310b049c5dcdcd931f046a3203313} Some additional changes to the original OpenZeppelin implementation where needed in order to make it compile with the current Solidity language (\texttt{solc} 5.0.0 and higher). We also changed the visibility of the \soltext{mint} function to public in order to be able to generate new tokens when needed in the simulation. The only changes we made to the other source files (\texttt{SafeMath.sol}, \texttt{IERC20.sol}) were and update of the compiler pragma to version 5.0.0.

All the necessary files including the compiled binaries\footnote{Compiled with the Remix IDE, \texttt{solc} 0.5.0+commit.1d4f565a with optimizations enabled.} can be found in the aforementioned Github repository.




\end{document}